\documentclass{article}

\PassOptionsToPackage{numbers, compress}{natbib}
    \usepackage[final]{neurips_2023_ml4ps}

\usepackage[utf8]{inputenc} 
\usepackage[T1]{fontenc}    
\usepackage{hyperref}       
\usepackage{url}            
\usepackage{booktabs}       
\usepackage{amsfonts}       
\usepackage{nicefrac}       
\usepackage{microtype}      
\usepackage{xcolor}         
\usepackage{multirow}
\usepackage{tabularx}
\usepackage{cleveref}
\usepackage{caption}
\usepackage{graphicx}
\usepackage{float}
\usepackage{subfigure}
\crefname{section}{§}{§§}

\title{E(2) Equivariant Neural Networks for Robust Galaxy Morphology Classification}

\makeatletter

\author{%
  Sneh Pandya\thanks{equal contribution} \thanks{correspondence to \texttt{pandya.sne@northeastern.edu}}\\
  Department of Physics\\
  Northeastern University\\
  NSF Institute for Artificial Intelligence and Fundamental Interactions  \\
  Boston, MA 02115 \\
\And
  Purvik Patel$^*$\\
  Khoury College of Computer Sciences \\
  Northeastern University \\
  Boston, MA 02115 \\
\And
    Franc O \\
    Khoury College of Computer Sciences \\
    Northeastern University \\
    Boston, MA 02115 \\
\And
    Jonathan Blazek \\
    Department of Physics \\
    Northeastern University \\
    Boston, MA 02115 \\
}

\begin{document}

\maketitle

\begin{abstract}
  We propose the use of group convolutional neural network architectures (GCNNs) equivariant to the 2D Euclidean group, $E(2)$, for the task of galaxy morphology classification by utilizing symmetries of the data present in galaxy images as an inductive bias in the architecture. We conduct robustness studies by introducing artificial perturbations via Poisson noise insertion and one-pixel adversarial attacks to simulate the effects of limited observational capabilities. We train, validate, and test GCNNs equivariant to discrete subgroups of $E(2)$ - the cyclic and dihedral groups of order $N$ - on the Galaxy10 DECals dataset and find that GCNNs achieve higher classification accuracy and are consistently more robust than their non-equivariant counterparts, with an architecture equivariant to the group $D_{16}$ achieving a $95.52 \pm 0.18\%$ test-set accuracy. We also find that the model loses $<6\%$ accuracy on a 50\%-noise dataset and all GCNNs are less susceptible to one-pixel perturbations than an identically constructed CNN. Our code is publicly available at \url{https://github.com/snehjp2/GCNNMorphology}.

\end{abstract}

\section{Introduction}
\label{introduction}

The study of galaxy morphology provides insight into how galaxies form and evolve over time. With the emergence of large-scale ground and space based observatories collecting massive amounts of data, deep learning has shown to be a promising candidate in its capabilities for powerful, efficient pattern recognition and inference across a variety of domains. Convolutional neural networks (CNNs) can be employed on the task of galaxy morphology classification \citep{Aniyan_2017, Lukic_2018, Hui_2022, scaife2021fanaroff} and be used to produce large galaxy morphological classification catalogues such as with the Dark Energy Survey Year 3 data \citep{Cheng_2021}. 

Group convolutional neural networks (GCNNs) \citep{groupcnn} utilize symmetries of the data as an inductive bias in the architecture by using a higher degree of weight sharing compared to typical CNNs. \citep{steerablecnns, weiler2019general} further introduced "Steerable CNNs", allowing fast implementation of GCNNs equivariant to $E(2)$. CNNs endowed with symmetry priors were used for galaxy morphology classification in \citep{dieleman2016exploiting, Dieleman_2015}.

The sensitivity of CNNs to adversarial perturbations has raised concerns about their practical deployment \citep{8294186}. Quite (in)famously, the output of CNNs can be drastically altered by perturbing one pixel in the input image \citep{Su_2019}, and there exists a variety of attacks invisible to human perception that can cause misbehavior in CNNs \citep{yuan2018adversarial}. This susceptibility to adversarial attacks has given rise to the study of robustness, wherein improvements in model training \citep{madry2019deep}, data preprocessing \citep{Xu_2018}, and architecture \citep{liu2018robust} have been studied. GCNNs robustness to geometric perturbations (e.g. rotations) were studied in \citep{dumont2018robustness}. \citep{DLastronomy, ciprijanovic2022deepadversaries} studied the effects of adversarial perturbations on deep learning algorithms in astronomy and the effectiveness of domain adaptation techniques in mitigating such attacks.

In this paper, we construct GCNNs utilizing the symmetries of the 2D Euclidean group, $E(2)$, which contains all rotations, translations, and reflections in flat space for the task of galaxy morphology classification, exploiting the fact that there is no canonical orientation for galaxies and inferring their morphological properties. $E(2)$ encompasses both the cyclic (rotations) and dihedral groups (rotations and reflections), $C_N$ and $D_N$, as discrete subgroups. We hypothesize that the robustness to symmetry transformations that are manifest in equivariant neural networks will result in increased robustness against adversarial perturbations that are relevant for the astronomical community. 

\section{Data \& Method}
\label{method}

\subsection{Galaxy10DECals Dataset}\label{section:data}
\label{dataset}

\begin{figure}
    \centering
    \includegraphics[width=\textwidth]{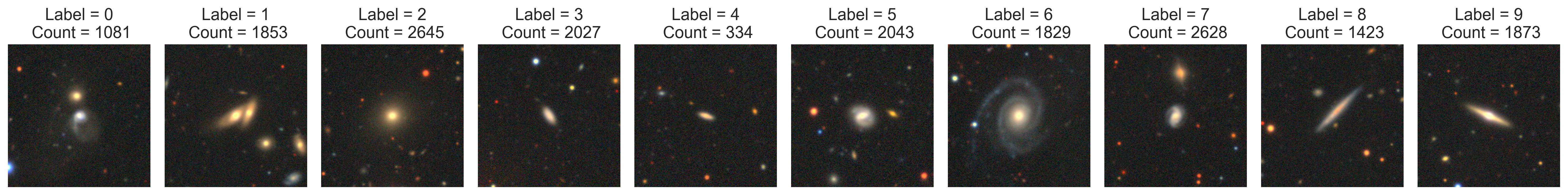}
    \caption{Examples of galaxy images and individual class abundances from the Galaxy10 DECals dataset containing $17,736$ labeled galaxies. Further information on each galaxy class can be found at \url{https://github.com/henrysky/Galaxy10}.}
    \label{fig:galaxy}
\end{figure}

The Galaxy10 DECals dataset \citep{2019MNRAS.483.3255L} contains $17,736$ colored galaxy images (in g, r, and z band) of 10 distinct, imbalanced classes and is comprised of data from the Galaxy Zoo \cite{Lintott_2008} Data Release 2 and DESI Legacy Imaging Surveys \cite{GZDR2, GZDecalsWalmsley, DESIDey} as shown in Figure \ref{fig:galaxy}. The galaxy labels are a result of multiple rounds of rigorous volunteer votes and filtering. Each galaxy image is further equipped cosmological redshift; the sample of galaxies used here are primarily low-redshift ($z \sim 0.1$).


We apply normalization to take image pixel values from $[0, 255]$ to $[-1, 1]$, and uniformly apply data augmentation comprised of random rotations, translations, and center-crops during training to ensure non-equivariant models do not have an inherent disadvantage in experiments compared to equivariant ones \citep{weiler2019general}. Internal experiments tested the effectiveness of morphological opening as used in \citep{Hui_2022} and the inclusion of spectroscopic redshift, but found no benefit. We create a 20\% test dataset of $3,542$ samples to evaluate the performance of our models, in which images are randomly rotated by an angle $\theta \in (0, 2\pi]$. We further conduct robustness studies by creating datasets with insertion of 25\%, 50\%, and 75\% Poisson noise -- as it closely resembles noise from CCD readouts or the atmosphere \citep{ciprijanovic2022deepadversaries} -- normalized with respect to the total original signal of the image. One-pixel attacks which simulate telescope processing errors were conducted using a differentiable genetic algorithm \citep{Su_2019}, in which a starting population of $400$ agents is evolved over $200$ generations to find pixel perturbations such that an image is misclassified by the model. One-pixel perturbations are a worst case scenario as a result of hardware failure at a detector level in an observational pipeline. These perturbations serve to test the out-of-distribution classification capabilities of our models in scenarios relevant to the astronomical community.

\subsection{Architectures}\label{section:architecture}



We utilize the software package \texttt{escnn} \citep{cesa2022a, e2cnn} to incorporate equivariance to all isometries of $E(2)$ \citep{weiler2019general} in constructing GCNNs equivariant to groups: $C_1$, $D_1$\footnote{We use the convention $D_N$ (as opposed to the more common $D_{2N}$) for the dihedral group, following \citep{weiler2019general}.}, $C_2$, $D_2$, $C_4$, $D_4$, $C_8$, $D_8$, $C_{16}$, and $D_{16}$ as described at the end of \cref{introduction}. Orders $N$ higher than 16 were not studied due to being too computationally intensive. These networks construct specialized filters that are themselves equivariant to the desired symmetries, and therefore learn transformation-invariant features that preserve the underlying symmetries in the data throughout training. In this way, we ensure generalization over such transformations and their persistence in the presence of perturbations.

We construct deep GCNNs with 11 convolutional blocks containing a group-convolutional layer, batch-normalization, and \texttt{ReLU} activation \citep{agarap2019deep}. Five intermittent pooling layers are used, which use point-wise average pooling for anti-aliased downsampling of feature maps. All features transform under the regular representation of the group, $\rho_{reg}$, with the hidden feature fields, $i$, including an expansion factor $f_i \in \{12, 24, 48, 48, 48, 48, 96, 96, 96, 112, 192\}$ to increase the number of convolutional channels. \texttt{GroupPooling} is used at the end to aggregate data across symmetry channels. The network lastly contains three fully-connected layers with \texttt{ELU} activation \citep{clevert2016fast} for classification. Internal experiments were conducted to find optimal network depth. The effectiveness of dropout \citep{dropout} and residual connections \citep{he2015deep} was also studied, and no substantial improvement in results was found. The CNN baseline with an identical architecture (with the exception of no \texttt{GroupPooling}) is constructed in \texttt{PyTorch} \citep{pytorch} utilizing the non-equivariant counterparts of components mentioned previously. Models that are equivariant to $SO(2)$ and $O(2)$ were constructed; however, GCNNs that are equivariant to continuous groups cannot transform under $\rho_{reg}$ and also require specific non-linearities \citep{franzen2021nonlinearities}. This prevents the creation of an identical CNN baseline for comparison to these models. As such, their performance is not presented here. 


\begin{figure}
    \centering
    \includegraphics[width=\textwidth]{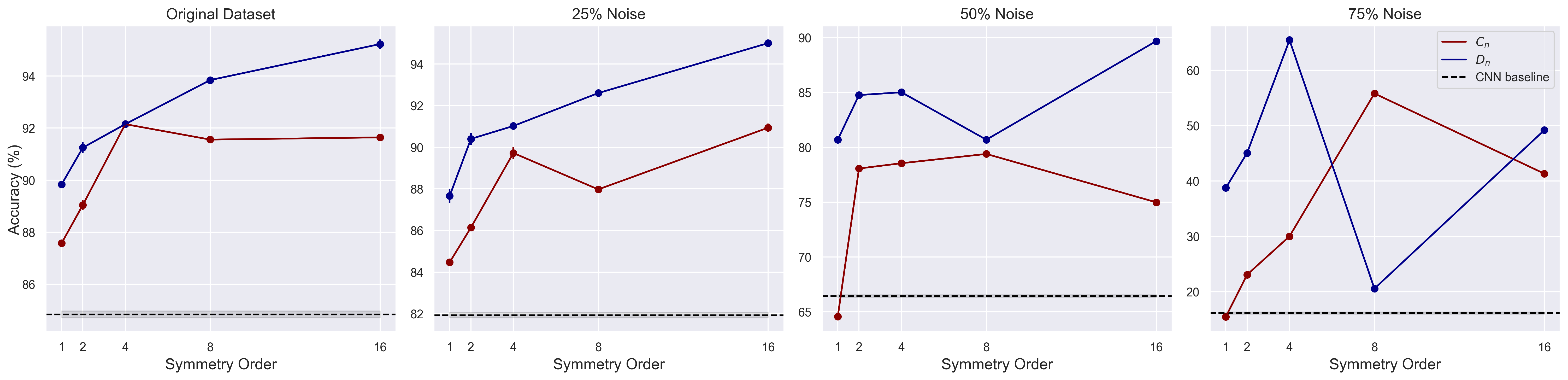}
    \caption{Model performance on test-set and noise experiments. $C_N$ results are in red, $D_N$ results in blue, and CNN baseline in black dashed-line. We see that equivariant models performance generally increases with $N$ and continually outperforms the CNN baseline.}
    \label{fig:results}
\end{figure}

\section{Results}

We trained all models using cross entropy loss, AdamW optimizer \citep{adamw}, and a step learning rate scheduler (10\% decay at 25-epoch intervals for a starting \texttt{lr} $=0.01$) for 100 epochs of training with early stopping on 4x NVIDIA A100-80GB GPUs. CNN training took $O$(2 hours), while GCNN training took anywhere from $O$(2 hours) to $O$(5 hours) for higher-order GCNNs. Despite having less parameters, the training of GCNNs is more costly than CNNs due to the added expense of group convolution. GCNNs are believed to converge much more rapidly than CNNs \citep{weiler2019general}, however this was not observed in our training. Classification uncertainties were obtained using bootstrap resampling.


\subsection{Noise Experiments}
\label{sec:noise_experiments}

On the original dataset, we find that all GCNNs outperform the CNN baseline, and the performance of GCNNs increases with the group order and inclusion of reflection symmetry. The $D_{16}$ model achieves a $95.22 \pm 0.18\%$ accuracy, compared to the CNN with $84.84 \pm 0.14\%$ accuracy and outperforming the DenseNet in \citep{Hui_2022} by 6\%. On the $25\%$ and $50\%$ noise datasets, the $D_{16}$ model achieves a $95.00 \pm 0.18\%$ and $89.67 \pm 0.16\%$ accuracy, compared to $81.93 \pm 0.13\%$ and $66.43 \pm 0.13\%$ for the CNN baseline. It is interesting to note that $C_{1}$ only contains the identity element, $\mathbf{e}$, and is therefore a trivial GCNN. Despite this, it still outperforms the CNN baseline in the original and 25\% noise dataset which suggests the equivariance constraint, however minimal, is still providing some benefit over traditional convolution.


On the 75\% noise dataset, we do not find strict performance benefits with increasing group order and symmetry. This can potentially be attributed to higher order GCNNs being more susceptible to symmetry breaking from the added perturbations; and as such the equivariance constraint of the model overpowers the expressivity as studied in \citep{wang2022approximately, finzi2021residual, canwehaveitall}. 
The highest performing GCNN is the $D_4$ with $65.47 \pm 0.28\%$ accuracy, substantially outperforming the CNN baseline with $16.12 \pm 0.17\%$. 

\begin{figure}
    \centering
    \includegraphics[width=\textwidth]{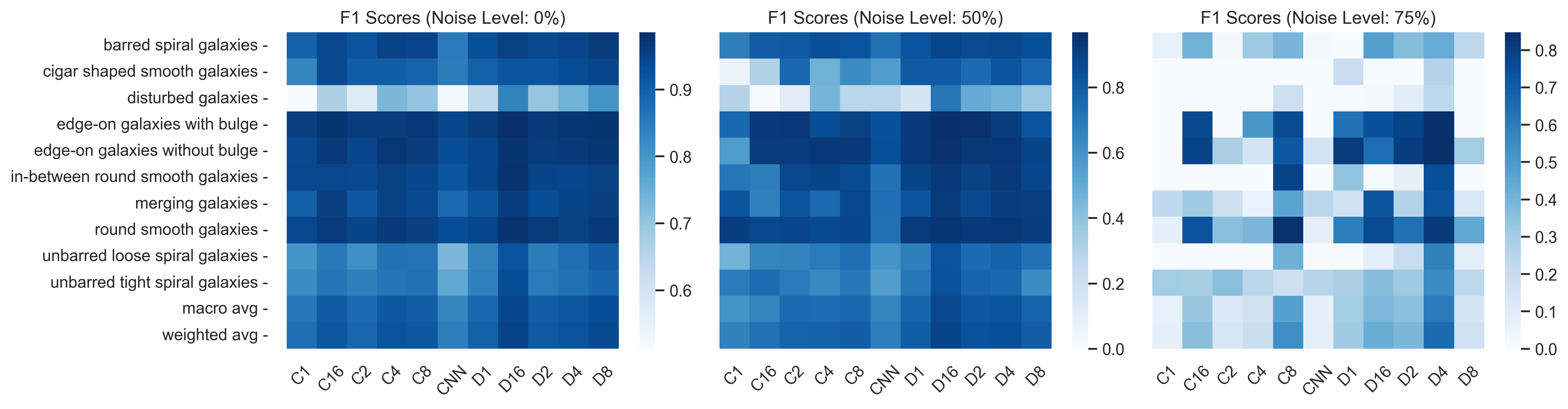}
    \caption{F1-score heatmaps for all models. We see that all models struggled to properly classify disturbed galaxies. It can be seen that in the presence of severe noise, the CNN struggles to properly classify all classes, while the best-performing $D_4$ model significantly struggles in three classes.}
    \label{fig:f1}
\end{figure}

As shown in Figure \ref{fig:f1}, disturbed galaxies were the most difficult to classify for all models, likely due to their high variability in morphology and lack of persistent structure. There was also confusion between unbarred tight and loose spiral galaxies, which are also easy to confuse upon human evaluation. The similarity of the weighted and macro average F1-scores indicates that the class imbalance of the dataset did not have a significant effect. 
Figure \ref{fig:f1} also illuminates that the success of the $D_4$ model in the $75\%$ noise dataset was in its ability to classify unbarred and in-between round-smooth galaxies, which other GCNNs struggled to classify.

\subsection{One-Pixel Attacks}

We perform one-pixel adversarial attack experiments to simulate the effects of data processing noise sustained by image compression or telescope errors following the work of \citep{ciprijanovic2022deepadversaries}.
We apply the differentiable genetic algorithm to each image in the test set of $3,542$ images for each model to see if a misclassifications can be made within 200 evolutions of the population as discussed in \cref{method}. The experiment was not able to be run on the $D_{16}$ model due to inadequate computational resources.
\vspace{-2mm}
\begin{table}[H]\label{tab:table}
\centering
\begin{tabular}{|l|l|l|l|l|l|l|l|l|l|l|l|}
\hline
\textbf{Model} & CNN  & $C_1$ & $C_2$ & $C_4$ & $C_8$ & $C_{16}$ & $D_1$ & $D_2$ & $D_4$ & $D_8$ & $D_{16}$ \\ \hline
\textbf{Incorrect (\%)} & 3.60 & 3.60  & 1.90  & 1.90  & 2.35  & 3.15     & 2.20  & 1.75  & 2.05  & 2.31  & N/A      \\ \hline
\end{tabular}
\vspace{-4mm}
\end{table}

The amount of images susceptible to one-pixel attacks from the genetic algorithm is less for GCNNs, however the previous expectation that GCNNs with higher order symmetries are more robust does not apply here as shown in the table. The lowest susceptibility was $1.75\%$ of test-set images for the $D_2$ model. Susceptibility to one-pixel attacks is empirically ubiquitous, and their manifestations in astronomical settings are completely random. The results here communicate that within a reasonable search-space for a stochastic algorithm, one-pixel perturbations that can misclassify GCNNs \emph{are more sparse and more difficult to find}, which suggest their increased robustness in a deployed setting.

\begin{figure}
    \centering
    \includegraphics[width=\textwidth]{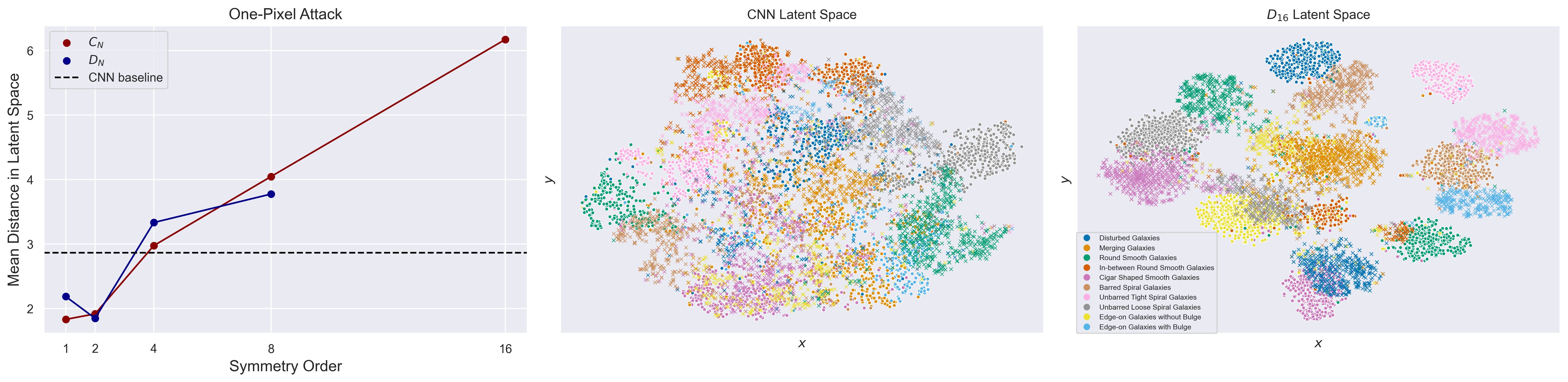}
    \caption{(Left) Mean Euclidean distance traversed in latent space for misclassified images in one-pixel experiments. (Middle / Right) t-SNE space constructions for CNN and $D_{16}$ models for original (circle marker) and 50\% noise (X marker) datasets. Latent space representations are the outputs of the model before the final fully connected layers.}
    \label{fig:latent}
\end{figure}

\section{Discussion \& Future Work}

We briefly analyze the models' learned latent space using t-SNE \citep{tsne}. As shown in Figure \ref{fig:latent} (left), for one-pixel experiments, the expectation that increased robustness implies that images travel farther in latent space  in order to cross a decision boundary (as seen in \citep{ciprijanovic2022deepadversaries}) is not observed, as the $D_{2}$ model exhibits a mean traversed (Euclidean) distance less than the CNN. However, the mean distance for $C_{16}$ - which is also more robust than the CNN - is twice as high as the CNN.  It's also seen in Figure \ref{fig:latent} (middle, right) that the $D_{16}$ can better cluster galaxy classes in the 50\% noise dataset and can therefore discern their properties better, but there is still significant movement and cluster-overlap in the latent space for some classes, which can explain the drop in accuracy seen in Figure \ref{fig:results}. The significance of these effects increase with the amount of noise added, and can be mitigated in networks trained with domain adaptation \citep{ciprijanovic2022deepadversaries}.
The CNN movement in latent space is overall less, however the decision boundaries are significantly more ill-defined which can explain the model's lack of robustness. In the future, it would be beneficial to conduct a more extensive analysis on the learned representations of GCNNs to interpret their increased robustness on this task.


Our work suggests that GCNNs are more robust than typical CNNs against adversarial perturbations that are common in astronomical imaging pipelines. Our models are able to perform with minimal data preprocessing to correct the class imbalance in our datasets or enhance the central galactic structure. In the future, the inclusion of domain adaptation in training and spectroscopic redshift as a feature may improve robustness. It would also be interesting to study robustness with respect to high redshift galaxies and dim low-redshift galaxies which are noisier and fainter than the sample of galaxies used here. The application of equivariant models has the potential to enhance the classification of over 20 billion galaxies from the Legacy Survey of Space and Time (LSST) at the Vera Rubin Observatory \citep{LSST}, deepening insights into their formation and evolution.


\section{Broader Impacts}

The techniques presented here have the potential to classify and extract features from images of arbitrary orientation and of significantly degraded quality, and as such warrant ethical concerns for maladaptations of this work. The exploitation of computer vision technologies for uses of surveillance is a poison. The authors steadfastly abhor the use of deep learning for purposes that do not seek to further scientific knowledge or provide a beneficial and equitable service to society.

\section{Acknowledgments and Disclosure of Funding}
We thank the anonymous referees for their useful comments. SP acknowledges support from the National Science Foundation under Cooperative Agreement PHY- 2019786 (The NSF AI Institute for Artificial Intelligence and Fundamental Interactions, \url{https://iaifi.org}. SP, PP, and FO acknowledge support from graduate teaching assistanships at Northeastern University. We thank Aleksandra Ćiprijanović at Fermilab for her useful comments on this work, and Robin Walters at Northeastern University for introducing us to equivariance. Computations were run on the FASRC Cannon cluster supported by the FAS Division of Science Research Computing Group at Harvard University.

\newpage
\bibliography{refs}

\newpage
\section{Appendix}
\begin{table}[h!]
    \centering
    \begin{tabular}{|ccccc|}
        \hline
        \textbf{Layers} & \textbf{Properties} & \textbf{Stride} & \textbf{Padding} & \textbf{Output Shape} \\
        \hline
        Input & 3 x 255 x 255 & & & \\
        \hline
        Conv2D & Filters: 12 & & &\\
        (w/ BatchNorm2D) & Kernel: 3x3 & 2 & 2 & (12, 129, 129) \\
        & Activation: ReLU & & &\\
        \hline
        Conv2D & Filters: 24 & & &\\
        (w/ BatchNorm2D)& Kernel: 3x3 & 1 & 1 & (24, 129, 129) \\
        & Activation: ReLU & & &\\
        \hline
        MaxPool2D & Kernel: 2 & 2  & 0 & (24, 64, 64) \\
        \hline
        Conv2D & Filters: 48 &  & &\\
        (w/ BatchNorm2D)& Kernel: 3x3 & 1 & 1 & (48, 64, 64)\\
        & Activation: ReLU & & &\\
        \hline
        Conv2D & Filters: 48 & & &\\
        (w/ BatchNorm2D)& Kernel: 3x3 & 1 & 1 & (48, 64, 64)\\
        & Activation: ReLU & & &\\
        \hline
        MaxPool2D & Kernel: 2 & 2 & 0 & (48, 32, 32)\\
        \hline
        Conv2D & Filters: 48 & & &\\
        (w/ BatchNorm2D)& Kernel: 3x3 & 1 & 1 & (48, 32, 32)\\
        & Activation: ReLU & & &\\
        \hline
        Conv2D & Filters: 48 & & &\\
        (w/ BatchNorm2D)& Kernel: 3x3 & 1 & 1 & (48, 32, 32)\\
        & Activation: ReLU & & &\\
        \hline
        Conv2D & Filters: 96 & & &\\
        (w/ BatchNorm2D)& Kernel: 3x3 & 1 & 1 & (96, 32, 32)\\
        & Activation: ReLU & & &\\
        \hline
        MaxPool2D & Kernel: 2 & 2 & 0 & (96, 16, 16)\\
        \hline
        Conv2D & Filters: 96 & & &\\
        (w/ BatchNorm2D)& Kernel: 3x3 & 1 & 1  & (96, 16, 16)\\
        & Activation: ReLU & & & \\
        \hline
        Conv2D & Filters: 96 & & &\\
        (w/ BatchNorm2D)& Kernel: 3x3 & 1 & 1 & (96, 16, 16) \\
        & Activation: ReLU & & & \\
        \hline
        MaxPool2D & Kernel: 2 & 2 & 0 & (96, 8, 8)\\
        \hline
        Conv2D & Filters: 112 & & & \\
        (w/ BatchNorm2D)& Kernel: 3x3 & 1 & 1 & (112, 8, 8) \\
        & Activation: ReLU & & & \\
        \hline
        Conv2D & Filters: 192 & & & \\
        (w/ BatchNorm2D)& Kernel: 3x3 & 1 & 1 & (192, 8, 8)\\
        & Activation: ReLU & & &\\
        \hline
        MaxPool2D & Kernel: 2 & 2 & 0 & (192, 4, 4) \\
        \hline
        Linear & Input Dimension: 3072 & & & \\
        (w/ BatchNorm1D) & Output Dimension: 64 & & & (, 64) \\
        & Activation: ELU  & & &\\
        \hline
        Linear & Input Dimension: 64 & & & \\
        (w/ BatchNorm1D) & Output Dimension: 32 & & & (, 32)\\
        & Activation: ELU & & & \\
        \hline
        Linear: & Input Dimension: 32& & & \\
         & Output Dimension: 10 & & & (, 10)\\
          & Activation: None & & &\\
        \hline
    \end{tabular}
    \caption{Architecture of CNN baseline.}
    \label{table:cnn_table}
\end{table}

\newpage
\begin{table}[h!]
    \centering
    \begin{tabular}{|ccccc|}
        \hline
        \textbf{Layers} & \textbf{Properties} & \textbf{Stride} & \textbf{Padding} & \textbf{Output Shape}  \\
        \hline
        Input & 3 x 255 x 255 & & &  \\
        \hline
        MaskModule & Margin: 1 & & & (3 x 255 x 255) \\
         & Sigma: 2.0  &  &  &   \\
        \hline
        R2Conv & Filters: 12 & & & \\
        (w/ InnerBatchNorm) & Kernel: 3x3 & 2 & 2 & ($N$ * 12, 129, 129)  \\
        & Activation: ReLU & & & \\
        \hline
        R2Conv & Filters: 24 & & & \\
        (w/ InnerBatchNorm)& Kernel: 3x3 & 1 & 1 & ($N$ * 24, 129, 129)  \\
        & Activation: ReLU & & & \\
        \hline
        PointwiseAvgPoolAntialiased & Sigma: 0.66 & 2  & 0 & ($N$ * 24, 65, 65)  \\
        \hline
        R2Conv & Filters: 48 &  & & \\
        (w/ InnerBatchNorm)& Kernel: 3x3 & 1 & 1 & ($N$ * 48, 65, 65)  \\
        & Activation: ReLU & & & \\
        \hline
        R2Conv & Filters: 48 & & & \\
        (w/ InnerBatchNorm)& Kernel: 3x3 & 1 & 1 & ($N$ * 48, 65, 65) \\
        & Activation: ReLU & & & \\
        \hline
        PointwiseAvgPoolAntialiased & Sigma: 0.66 & 2 & 0 & ($N$ * 48, 33, 33) \\
        \hline
        R2Conv & Filters: 48 & & & \\
        (w/ InnerBatchNorm)& Kernel: 3x3 & 1 & 1 & ($N$ * 48, 33, 33)\\
        & Activation: ReLU & & & \\
        \hline
        R2Conv & Filters: 48 & & & \\
        (w/ InnerBatchNorm)& Kernel: 3x3 & 1 & 1 & ($N$ * 48, 33, 33) \\
        & Activation: ReLU & & & \\
        \hline
        R2Conv & Filters: 96 & & & \\
        (w/ InnerBatchNorm)& Kernel: 3x3 & 1 & 1 & ($N$ * 96, 33, 33)  \\
        & Activation: ReLU & & & \\
        \hline
        PointwiseAvgPoolAntialiased & Sigma: 0.66 & 2 & 0 & ($N$ * 96, 17, 17) \\
        \hline
        R2Conv & Filters: 96 & & & \\
        (w/ InnerBatchNorm)& Kernel: 3x3 & 1 & 1  & ($N$ * 96, 17, 17) \\
        & Activation: ReLU & & & \\
        \hline
        R2Conv & Filters: 96 & & & \\
        (w/ InnerBatchNorm)& Kernel: 3x3 & 1 & 1 & ($N$ * 96, 17, 17) \\
        & Activation: ReLU & & & \\
        \hline
        PointwiseAvgPoolAntialiased & Sigma: 0.66 & 2 & 0 & ($N$ * 96, 9, 9) \\
        \hline
        R2Conv & Filters: 112 & & & \\
        (w/ InnerBatchNorm)& Kernel: 3x3 & 1 & 1 & ($N$ * 112, 9, 9)  \\
        & Activation: ReLU & & &\\
        \hline
        R2Conv & Filters: 192 & & & \\
        (w/ InnerBatchNorm)& Kernel: 3x3 & 1 & 1 & ($N$ * 192, 9, 9) \\
        & Activation: ReLU & & & \\
        \hline
        PointwiseAvgPoolAntialiased & Sigma: 0.66 & 2 & 0 & ($N$ * 192, 5, 5)  \\
        \hline
        GroupPooling &  &  &  & (192, 5, 5)  \\
        \hline
        Linear & Input Dimension: 4800 & & &  \\
        (w/ BatchNorm1D) & Output Dimension: 64 & & & (, 64)  \\
        & Activation: ELU  & & & \\
        \hline
        Linear & Input Dimension: 64 & & & \\
        (w/ BatchNorm1D) & Output Dimension: 32 & & & (, 32)  \\
        & Activation: ELU & & & \\
        \hline
        Linear: & Input Dimension: 32& & & \\
         & Output Dimension: 10 & & & (, 10) \\
          & Activation: None & & & \\
        \hline
    \end{tabular}
    \caption{Architecture of GCNNs. In constructing GCNNs equivariant to $C_N$, $N$ in the output shape is exactly equal to the group order. For GCNNs equivariant to $D_N$, let $N \rightarrow 2N$.}
    \label{table:gcnn_table}
\end{table}

\end{document}